\documentclass[conference]{IEEEtran}
\IEEEoverridecommandlockouts
\usepackage{cite}
\usepackage{amsmath,amssymb,amsfonts}
\usepackage{algorithmic}
\usepackage{graphicx}
\usepackage{textcomp}
\usepackage{xcolor}
\def\BibTeX{{\rm B\kern-.05em{\sc i\kern-.025em b}\kern-.08em
    T\kern-.1667em\lower.7ex\hbox{E}\kern-.125emX}}
\begin{document}

\title{Autonomous Choreography of WebAssembly Workloads in the Federated Cloud-Edge-IoT Continuum
}

\author{\IEEEauthorblockN{Piotr Sowiński}
\IEEEauthorblockA{\textit{Warsaw University of Technology} \\
Warsaw, Poland \\
0000-0002-2543-9461}
\and
\IEEEauthorblockN{Ignacio Lacalle}
\IEEEauthorblockA{\textit{Communications Department} \\
\textit{Universitat Politècnica de València}\\
Valencia, Spain \\
0000-0002-6002-4050}
\and
\IEEEauthorblockN{Rafael Vaño}
\IEEEauthorblockA{\textit{Communications Department} \\
\textit{Universitat Politècnica de València}\\
Valencia, Spain \\
0000-0003-2372-6253}

\and
\IEEEauthorblockN{Carlos E. Palau}
\IEEEauthorblockA{\textit{Communications Department} \\
\textit{Universitat Politècnica de València}\\
Valencia, Spain \\
0000-0002-3795-5404}
}

\maketitle

\begin{abstract}
The concept of the federated Cloud-Edge-IoT continuum promises to alleviate many woes of current systems, improving resource use, energy efficiency, quality of service, and more. However, this continuum is still far from being realized in practice, with no comprehensive solutions for developing, deploying, and managing continuum-native applications. Breakthrough innovations and novel system architectures are needed to cope with the ever-increasing heterogeneity and the multi-stakeholder nature of computing resources. This work proposes a novel architecture for choreographing workloads in the continuum, attempting to address these challenges. The architecture that tackles this issue comprehensively, spanning from the workloads themselves, through networking and data exchange, up to the orchestration and choreography mechanisms. The concept emphasizes the use of varied AI techniques, enabling autonomous and intelligent management of resources and workloads. Open standards are also a key part of the proposition, making it possible to fully engage third parties in multi-stakeholder scenarios. Although the presented architecture is promising, much work is required to realize it in practice. To this end, the key directions for future research are outlined.
\end{abstract}


\section{Introduction}
The concept of the computing continuum has gained popularity in the last few years~\cite{continuum}. It embraces the idea of assembling a wide variety of heterogeneous computing resources as a single manageable entity (spanning IoT devices, edge nodes, private or public clouds, etc.). Technologically, the continuum approach should achieve better management of the widespread resources to simplify the execution of workloads (applications, services, workflows…) leveraging aspects such as network virtualization, energy management, performance, dynamic demand by on-going services, and more. This would have many benefits, including increased efficiency, longer useful lifespan of computing equipment, effective collaboration among stakeholders, new use cases, and improved privacy and sovereignty, among others. On top, the fact of addressing all the previous based on intelligence/AI mechanisms, is what derives the so-called cognitive computing continuum~\cite{ferrer2021towards}.

However, nowadays, such continuum is far from being seamlessly realized in practice. It appears complicated for code developers, system owners, and end users alike. Underlying heterogeneity in such environments remains unresolved, with devices running different operating systems and having varied computing capabilities. Existing technologies do not fit together seamlessly or do not cover all complexities, requiring deep knowledge and expertise to barely grasp their surface. Also, applications depend heavily on the underlying libraries, CPU architectures, and resource management frameworks. As long as there is a lack of \emph{continuum-native} tools and mechanisms, organizations will struggle to use it, due to the high associated costs and risks. Furthermore, current applications are unnecessarily heavy, due to bloated container images and ineffective use of available resources, resulting in high energy and infrastructure costs.

Yet, overcoming these realities seems quite challenging. First, whenever a service needs to work with data, deep knowledge of data formats and models is required, hindering fast and actionable use cases to take place dynamically. Second, current mechanisms and solutions are not prepared for scalability and adaptability, requiring deep and complex updates whenever new innovations emerge. Third, seamless cooperation among nodes is still unrealized, as only a few of them can autonomously apply AI capabilities in a controlled manner, while the continuum does not stretch well to the vast majority of them. Their variety (unaddressed from an abstraction perspective) prevents the formation of a truly decentralized swarm. This challenge is very tough due to varied operating systems and their restrictions, limited compute capabilities, and diverse compilation mechanisms.

This work proposes an architecture for autonomous choreography (advanced form of orchestration) of workloads, network, and data in a federated cognitive continuum. A vision is presented in which the orchestration of containerized and WebAssembly workloads is managed autonomously by AI-driven schedulers, enabling smooth choreography in nodes across the continuum, optimizing parameters, such as energy use, carbon footprint, or QoS.

The rest of the paper is organized as follows: Section~\ref{sec:background} outlines the background of the various technical fields involved, understanding the current challenges and on-going initiatives. Section~\ref{sec:architecture} describes the proposed architecture, how the different ideas glue together and a theoretical implementation approach. 
Finally, Section~\ref{sec:conclusion} draws early conclusions on the viability and usefulness of the architecture and envisions the next steps to have it materialized.  

\section{Background}
\label{sec:background}

Coping with heterogeneous computing resources and frameworks when creating applications is not a new field of research. For instance, developers around the globe have widely made use of the Java Virtual Machine (JVM)~\cite{java} as an execution environment that can be used in the continuum, supporting servers and some embedded devices. However, JVM's programming language support is limited, and it has a significant performance overhead. On another note, containers are a very good tool for abstracting underlying resources and delivering generic applications~\cite{containers}. However, they can still be large, do not support the smallest of devices, and are built for specific architectures and operating systems. The emerging WebAssembly (Wasm) promises to enable truly portable software that can execute with minimal performance losses and small memory footprints, even on IoT devices. The Wasm ecosystem is developing rapidly~\cite{vano2023cloud}, with initiatives like WASI\footnote{https://wasi.dev/} (portable system interface) and WAMR\footnote{https://github.com/bytecodealliance/wasm-micro-runtime} (lightweight and portable runtime). However, Wasm is a very new technology and its potential in the continuum is yet to be realized~\cite{menetrey2022webassembly}.

Another relevant field for this work is the orchestrated deployment of workloads in a computing continuum across various environments and stakeholders. Existing cloud and edge paradigms are mainly monolithic, siloed, and often constrained to a single vendor’s ecosystem. Here, cloud providers are suggesting new ways of multi-clustering strategies~\cite{ASENSIO2021157}. Also, serverless solutions (that address scaling, boot time, and over-management concerns) came out with Severless4IoT~\cite{serverless4iot} and OpenWolf\cite{sicari}, among others. However, they either do not cover IoT devices, or are limited in the types of workloads they can manage. New platforms are now emerging with enhanced capabilities to support edge clusters, such as RedHat OpenShift\footnote{https://www.redhat.com/en/technologies/cloud-computing/openshift}, OKD\footnote{https://okd.io/}, Kubernetes-based deployments (k0s\footnote{https://k0sproject.io/} and K3s\footnote{https://k3s.io/}), and multi-cluster management such as Nuvla/NuvlaBox\footnote{https://sixsq.com/}, OCM\footnote{https://open-cluster-management.io/}, Fleet\footnote{https://fleet.rancher.io/}, and LIQO\footnote{https://docs.liqo.io/en/v0.7.2/}. In the context of cloud computing, related work also considers the need for operating systems specifically designed to dynamically manage datacenter resources, with some software solutions already in existence, such as Apache Mesos\footnote{https://mesos.apache.org/} or Mesosphere DC/OS\footnote{https://dcos.io/}. All in all, there is currently no available solution for supporting hyper-distributed, heterogeneous, collaborative systems able to deploy services in IoT devices, edge nodes, and cloud providers alike, spanning multiple management domains (hybrid cloud).

The capacity to dynamically orchestrate such deployed workloads is also pursued in the literature. It seems not longer possible to optimize energy efficiency, cost, carbon footprint, and other factors manually or with a rigid algorithm deciding where to deploy workloads~\cite{kokkonen2023autonomy}. In a multi-stakeholder, federated environment assuming hierarchical control over every node in the continuum is unfeasible, therefore, autonomy is needed~\cite{pujol}. Thus, recent works indicate that flexible, robust, and intelligent solutions are needed that can promptly and autonomously manage workloads. Here is where AI-driven workload scheduling emerges as a very active research area, but with practical implementations still in their infancy. Narrow formulations of the task were proposed, based on the Function as a Service (FaaS) paradigm~\cite{fatouros}. Otherwise, small lab demonstrators~\cite{divide} or mostly theoretical proposals of dynamic scheduling of workloads to optimize energy are the main explored topics~\cite{llima}. However, for the solution to orchestration and deployment to be applicable in a truly federated environment, it must be based on open standards and allow anyone to bring their own scheduler implementation that would still be compatible with the rest of the continuum. These capabilities are currently not present in state of the art research. 

Managing data in a unified way across the continuum is also an open research topic. Diverse data types, formats, computing capabilities or the usage of diverse tools exacerbate the complexity, asking developers to have extensive knowledge of the underlying data sources, formats, APIs, permissions, reliability, and other details. The application must also be aware of the network environment, which can vary greatly between platforms. Commercial solutions for creating a ``data fabric'' exist, but focus on vendor-locked, cloud-only environments (e.g., IBM, K2View, Talend). In the literature, MEDAL~\cite{medal} is a promising concept with no implementation yet, where data applications are managed with ``Data Fibers'', however, it does not employ AI. Some research projects such as RE4DY\footnote{https://re4dy.eu/} and aerOS\footnote{https://aeros-project.eu} propose ``Virtual Data Containers'' and ``Data Fabric'' concepts, that are, by now, only theoretical and do not promise to support hyper-distributed, multi-stakeholder systems with all outlined requirements. Eclipse Zenoh\footnote{https://newsroom.eclipse.org/eclipse-newsletter/2021/july/eclipse-zenoh-edge-data-fabric} is a relevant open-source project with a data management suite that includes its own networking layer that was tested in IoT use cases~\cite{zenoh}. Regarding network abstraction for data services, PuzzleMesh~\cite{puzzle} and SDFog-Mesh~\cite{sdfogmesh} can be found in the literature. Most significant advances in this aspect are in the field of cloud-native open solutions like Open Service Mesh\footnote{https://www.cncf.io/projects/open-service-mesh/}, Calico\footnote{https://www.cncf.io/online-programs/calico-networking-with-ebpf/}, Cilium\footnote{https://cilium.io/} or flannel\footnote{https://github.com/flannel-io/flannel}, but those are limited to specific container frameworks (K8s), and do not care about data, privacy, or governance concerns. There is currently no solution that supports data management and sovereignty in the continuum acknowledging the privacy, networking, and holistic coverage needs.

Lastly, growing attention is also being paid to making the continuum more secure and actionable by developers and users. Developing a cloud application is radically different to IoT and everything in between, as are the execution environments, available software, network stacks, hardware capabilities, and developer tools. There exist several attempts at solving these problems, with the most prominent trend being to apply the highly successful cloud-native computing principles to the edge. However, this approach cannot scale to the smallest of devices due to hardware limitations. To counter this issue, the use of WebAssembly throughout the continuum is often cited as the most promising solution, but the technology is still in its early stages~\cite{vano2023cloud}. Realizing continuous integration / continuous deployment (CI/CD) in such a continuum is similarly challenging, with current solutions limited to only a small part of it (such as the plentiful cloud-oriented DevOps solutions~\cite{Leite_2019} or Renode\footnote{https://renode.io/} focusing on IoT devices), or function only within a single closed ecosystem (e.g., Azure DevOps\footnote{https://learn.microsoft.com/en-us/azure/iot-edge/how-to-continuous-integration-continuous-deployment-classic?view=iotedge-1.4} for Azure IoT Edge). Security-wise, a promising trend are trusted execution environments (TEE) that isolate code execution in hardware. However, the different proprietary TEE solutions pose portability challenges. Scontain~\cite{scontain} and Azure Sphere\footnote{https://azure.microsoft.com/en-us/products/azure-sphere/} have proposed using confidential containers, but these solutions are platform-specific and can be resource-intensive.

Ideally, there should be a single streamlined execution environment with consistent interfaces enabling the same code to be run on any machine in the cognitive computing continuum, and that is what this work aims at proposing.

\section{Proposed Architecture}
\label{sec:architecture}

\begin{figure}[tb]
    \centering
    \includegraphics[width=\columnwidth]{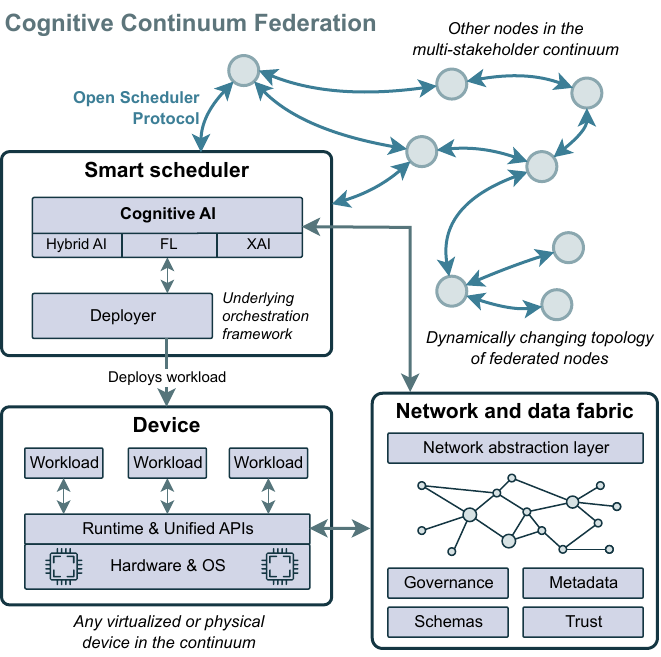}
    \caption{Overview of the proposed architecture.}
    \label{fig:overview}
\end{figure}

In this section a novel architecture is proposed, one that can comprehensively tackle the aforementioned challenges. The base idea for the architecture is ``any code, anywhere'', where computational workloads can be flexibly and intelligently scheduled on almost every device in the Cloud-Edge-IoT continuum, including the usually neglected resource-constrained devices. Figure~\ref{fig:overview} presents an overview of the proposed architecture.

The base building block of applications in the proposed concept is the unified compute module -- a platform-agnostic software package, that can run anywhere in the federated continuum. The modules can either utilize ``classic'' containerization, or the more portable, secure, and lightweight alternative of WebAssembly modules. The modules are then deployed in sandboxed environments using a variety of compatible runtimes. The continuum, naturally, consists of a wide variety of device types and execution environments, ranging from tiny IoT devices up to hyperscale data centres, and multiple CPU architectures (x86, Arm, RISC-V)~\cite{kimovski2021cloud, menetrey2022webassembly}. To efficiently manage and scale such overwhelmingly heterogeneous deployments, the architecture employs a decentralized approach to workload choreography and scheduling, using autonomous smart schedulers.

The schedulers are responsible for managing the workloads in their domain (e.g., a data centre, IoT deployment, an edge server) and use state-of-the-art AI techniques to optimize energy use, QoS, latency, etc. The schedulers can use AI techniques such as reinforcement learning to effectively adapt to environments where access to resources is determined by sophisticated and unpredictable factors. Hybrid AI techniques can also be used to fuse machine learning with semantic, robust knowledge about the environment~\cite{marcus2020next, garcez2023neurosymbolic}. The schedulers choreograph the workloads and resources (compute, storage, sensors…) in their domain, assigning workloads to specific compute resources while ensuring the consistency of the application. A scheduler can either run the workload in its domain or offload it to peer nodes. The schedulers act and communicate using the open scheduler protocol, which enables third-party implementations for any current and future platforms. The nodes are in a network with a dynamically changing topology (hierarchical or peer-to-peer), forming the Cognitive Continuum Federation -- capable of adapting its structure to the changing requirements and available resources. 

Communication between the workloads is handled by the end-to-end network and data fabric, enabling effortless communication with the rest of the system. The fabric abstracts away the underlying network and data exchange mechanisms. Here, the unified API of the unified compute modules plays a crucial role, providing the developer with a consistent interface, that works the same in the entire continuum. The data fabric has built-in provenance and active metadata tracking capabilities to cater for the requirements of data spaces.

\subsection{Unified Compute Modules}

\begin{figure}[tb]
    \centering
    \includegraphics[width=8cm]{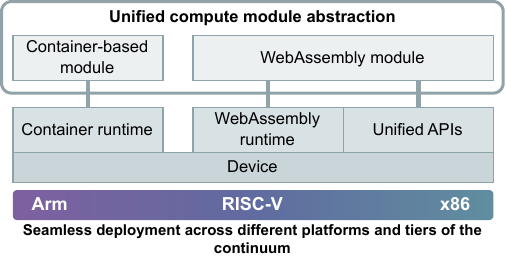}
    \caption{The unified compute module concept.}
    \label{fig:modules}
\end{figure}

The proposition for composing and deploying workloads across the continuum revolves around the idea of the unified compute module (Fig.~\ref{fig:modules}) -- a software package that, in principle, can run on any device and platform in the continuum, irrespective of the CPU’s architecture, or the operating system. The key technology behind this innovation is WebAssembly, a lightweight, universal binary format for application code that is then executed within isolated runtimes, giving much better security guarantees out-of-the-box, as compared to traditional containers. Two workload packaging formats are supported: WebAssembly and Open Container Initiative (OCI) containers\footnote{https://opencontainers.org/} (e.g., Docker), catering for a wider range of workloads. While WebAssembly is a much lighter paradigm, it is also limited in terms of supported interfaces and capabilities, and thus not all workloads can be easily converted to it, hence the need for containers. On the other hand, WebAssembly is improving rapidly and can support a much wider range of platforms~\cite{menetrey2022webassembly, wang2021empowering, spies2021evaluation, vano2023cloud}.

The unified API gives the applications robust access to the rest of the platform's capabilities (e.g., the network and data fabric), regardless of where in the continuum will the workload be deployed. The unified API should be ported or given bindings for several programming languages, to make it applicable as many applications as possible.

\subsection{Network and Data Fabric}

\begin{figure}[tb]
    \centering
    \includegraphics[width=8cm]{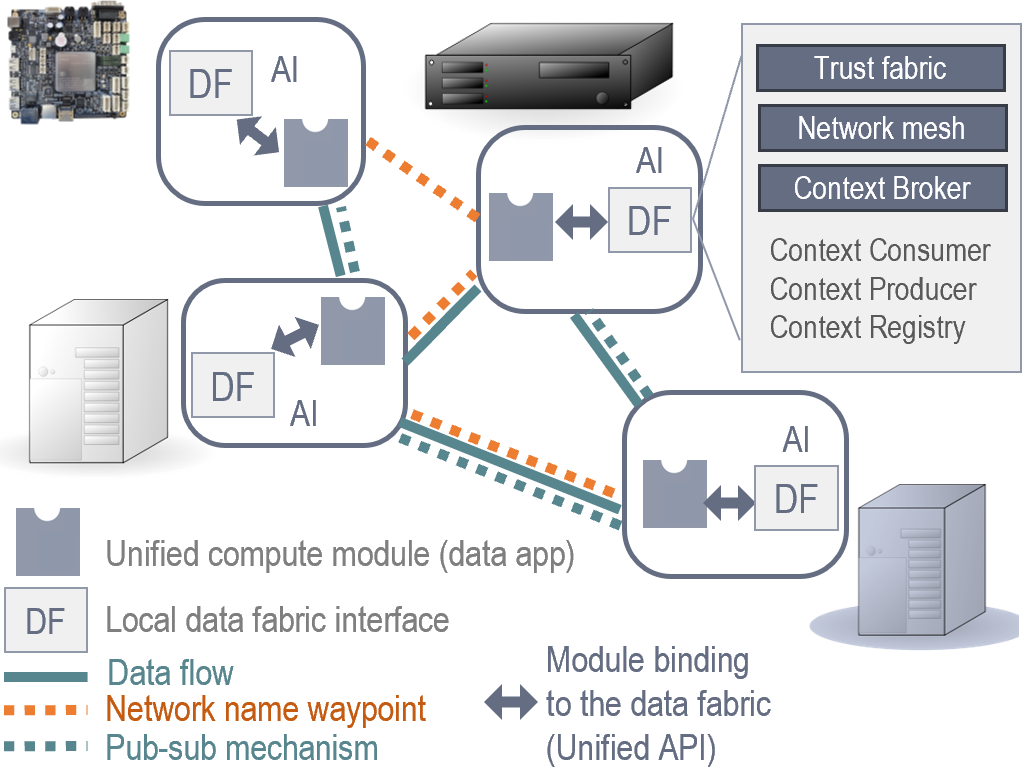}
    \caption{The network and data fabric.}
    \label{fig:fabric}
\end{figure}

The end-to-end network and data fabric (Fig.~\ref{fig:fabric}) is to be capable of connecting data services, abstracting the underlying complexity of trust, sovereignty, network connectivity, data types, and formats. To achieve the latter, the nodes in the fabric shall seamlessly exchange representations of metadata, data producers, and data consumers, in the form of knowledge graphs, using portable protocols (e.g., Jelly~\cite{sowinski2022efficient}). The graphs will be explicitly aligned with active metadata (description of the data) that will be governed by flexible rules and standards (e.g., NGSI-LD API). Active metadata implies that self-description of data will be facilitated by AI mechanisms, ontologies, and annotations, introducing automated data integration. Achieving seamless knowledge sharing between nodes (both peer-to-peer and in hierarchical domains) will allow producers and consumers (services) to be decoupled from specific formats, types, conversions, and topologies. In addition, data exchange will follow publish/subscribe mechanisms (similar to the MQTT protocol) supporting reliability, dynamic discovery, data cataloguing, and ownership. The Context Broker (assisted by intermediate data aggregation) will expose data to consumers. This will also allow to deploy the data spaces principles in the data fabric. Data will, by default, remain in every node’s scope if this is what the application requires. Shared metadata will also be used to facilitate connections between consumers and producers’ data storage facilities. The data fabric will then enable on-demand data retrieval based on policies and security credentials. On another note, active metadata will be connected to the governance solution in the data fabric, including data source and endpoint mapping to security credentials and privacy labels. This directly links with another feature included in the data fabric: trust and governance. A self-sovereign identity management structure will be channeled through federated IdMs that can reside on different domains and be owned by different stakeholders.

The end-to-end network fabric for data services is also be needed to realize the presented vision. Inspiration can be drawn from current trends such as eBPF\footnote{https://ebpf.io/} connectivity of services (tied to K8s deployments) and the Istio ambient mesh\footnote{https://istio.io/v1.15/blog/2022/introducing-ambient-mesh/}. The proposed concept relies on the creation of secure tunnels (waypoints) connecting data services among themselves, based on service names. The network fabric can also include the automated management of pipelines, drawing from previous concepts and potential transfer of Apache Streamipes\footnote{https://streampipes.apache.org/} or Kiali’s approaches\footnote{https://kiali.io/}. Together, the network and data fabric will enable data plan management across the continuum, focusing on simplified, automated operations, compatibility, agility, and reduced costs.

\subsection{Smart Schedulers}

\begin{figure}[tb]
    \centering
    \includegraphics[width=8cm]{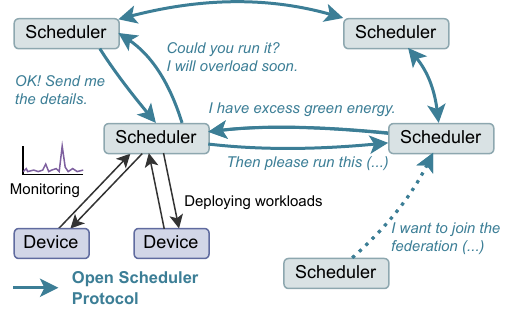}
    \caption{Smart schedulers cooperating via the open scheduler protocol.}
    \label{fig:schedulers}
\end{figure}

The proposed architecture aims to choreograph the resources of continuum ecosystems formed by the cloud, edge computing infrastructures (far, near), and IoT devices with different CPU architectures (including RISC-V), as well as operating systems (Fig.~\ref{fig:schedulers}). Here, choreography diverges from classic orchestration, as modern applications should not only rely on a central orchestrator to be deployed and to function~\cite{de2023relating}. They must have the capacity to act independently and thus be able to better adapt to the changing resources and requirements. The technical proposal here is two-fold: (1) to establish a federation choreography framework that will manage resource and service descriptions along with their proposed allocation, and (2) to rely on open, smart schedulers living on each node that will be responsible for deciding whether to run or offload workloads, in a decentralized way.

Through the first, the architecture will handle complex applications as workflows of stateless and stateful workloads, aiming to meet user-defined KPIs. It will implement an abstraction layer to describe the characteristics of custom applications composed of unified compute modules. Similarly, all available resources belonging to the continuum and their characteristics (including current performance, availability, trustworthiness, etc.) will be registered together with the data sources (e.g., sensors). This continuously updated registry will be dynamic and distributed among the nodes so that the federated entities (nodes) will have information about each other (relaxing the need for centralization). Regarding deployment, the framework will propose an initial deployment based on policy-like KPIs tailored to the application (eco-efficiency, performance, reliability, quality of service, etc.), aligning application’s requirements with available resources. AI will make the framework cognitive in three ways: (i) by interpreting and predictively managing KPIs of the applications, (ii) by optimizing workload distribution based on KPIs, prior knowledge, and heuristics, and (iii) by predicting application behaviour based on trends to make adjustments pre-emptively. At runtime, all monitored data will be used by the choreography framework, through AI, to learn the relationship between the behaviour of applications and their resources, and between application’s characteristics and its KPIs. The AI then will issue recommendations and imperative requests to adjust the deployment.

The second part of the technical proposition starts with designing a protocol for the open schedulers that will define the types of interactions the schedulers can engage in. The protocol will make few assumptions on the topology and authority structure, focusing on formulating an open specification that developers (including third parties) may use to build their own implementations. The schedulers will be able to use any AI methods, classical or machine learning (from deep neural networks to rule engines), to decide whether to take over the workload or to offload it to a peer within the federation. The open and decentralized-by-design protocol will facilitate the development of novel AI methods for continuum governance. To help accelerate adoption, a compatibility test kit will be provided, that will check if a given scheduler implementation is compatible with the specification. The creation of the protocol should follow a process in which the broader community (research and industry is included), eliciting crucial feedback.

\section{Conclusion and Future Work}
\label{sec:conclusion}

The proposed architecture attempts to tackle the very relevant problem of fully exploiting the potential of the federated Cloud-Edge-IoT continuum. The ever-increasing heterogeneity of computational resources (emerging CPU architectures, different operating systems, networking capabilities, etc.) require a fundamental change in how modern applications are designed. In the presented concept, this is embodied in the use of lightweight and portable WebAssembly modules, which can be easily deployed anywhere in the continuum. This flexibility is enabled in part by the unified API for modules. In this setting, the applications communicate via the cross-tier network and data fabric, greatly simplifying building the application, data governance, schema management, and more. Finally, to allow the workloads to be scheduled and adjusted, the smart schedulers are proposed, along with the open scheduler protocol. The protocol and the schedulers are what enables the vision of truly decentralized, autonomous choreography in the federated continuum.

Although the concept is promising, a significant amount of work is required to fully realize it. The unified compute module and unified API vision is already partially realized, with community's work underway on integrating WebAssembly with orchestration frameworks, container runtimes, and standardized interfaces such as WASI. However, these efforts are still at an early stage. For the network and data fabric, several of the required components were already demonstrated in some form in past research. However, the solutions are not integrated yet and largely not production-ready. For the smart schedulers, although a lot of research was published on the subject of AI-driven orchestration, few of these designs were implemented in practice. More work is especially required to standardize the scheduler's interfaces, aiming for an approach such as the proposed open scheduler protocol. In all of these efforts, the practice of open standards-first, implementation-second should be followed, to ensure the developed solutions are modular and can be extended by third-parties to suit their needs.

\bibliographystyle{IEEEtran}
\bibliography{bibliography}

\end{document}